\UseRawInputEncoding
\documentclass[%
 reprint,
superscriptaddress,
 amsmath,amssymb,
 aps,
]{revtex4-2}
\usepackage{graphicx} 
\usepackage{tikz}
\usepackage{natbib}%
\usepackage{hyperref}
\usepackage{subfigure}
\usepackage{amsmath}
\usepackage{comment}

\usepackage{svg}
\usepackage{booktabs}
\usepackage{multirow}
\usepackage{array}
\usepackage{makecell}

\begin{document}

\title{The Transmission Line Model for 2D Materials and van der Waals Heterostructures}

\author{Tomer Eini}
\affiliation{School of Electrical Engineering, Faculty of Engineering, Tel Aviv University, Tel Aviv 6997801, Israel}
\author{Anabel Atash Kahlon}
\affiliation{School of Electrical Engineering, Faculty of Engineering, Tel Aviv University, Tel Aviv 6997801, Israel}
\author{Matan Meshulam}
\affiliation{School of Electrical Engineering, Faculty of Engineering, Tel Aviv University, Tel Aviv 6997801, Israel}
\affiliation{School of Physics and Astronomy, Tel Aviv University, Tel Aviv, 6997801, Israel}
\author{Thomas Poirier}
\affiliation{Tim Taylor Department of Chemical Engineering, Kansas State University, Manhattan, KS, USA}
\author{James H. Edgar}
\affiliation{Tim Taylor Department of Chemical Engineering, Kansas State University, Manhattan, KS, USA}
\author{Seth Ariel Tongay}
\affiliation{Materials Science and Engineering, School for Engineering of Matter, Transport and Energy,
Arizona State University, Tempe, Arizona 85287, USA}
\author{Yarden Mazor}
\affiliation{School of Electrical Engineering, Faculty of Engineering, Tel Aviv University, Tel Aviv 6997801, Israel}
\author{Itai Epstein}
\email{itaieps@tauex.tau.ac.il}
\affiliation{School of Electrical Engineering, Faculty of Engineering, Tel Aviv University, Tel Aviv 6997801, Israel}
\affiliation{Center for Light-Matter Interaction, Tel Aviv University, Tel Aviv 6997801, Israel}
\affiliation{QuanTAU, Quantum Science and Technology Center, Tel Aviv University, Tel Aviv 6997801, Israel}

\date{July 2025}

\begin{abstract}
Van der Waals heterostructures (VdWHs) composed of stacked two-dimensional (2D) materials have attracted significant attention in recent years due to their intriguing optical properties, such as strong light-matter interactions and large intrinsic anisotropy. In particular, VdWHs support a variety of polaritons—hybrid quasiparticles arising from the coupling between electromagnetic waves and material excitations—enabling the confinement of electromagnetic radiation to atomic scales. The ability to predict and simulate the optical response of 2D materials heterostructures is thus of high importance, being commonly performed until now via methods such as the transfer-matrix-method, or Fresnel equations. While straight forward, for complex structures these often yield long and complicated expressions, limiting intuitive and simple access to the underlying physical mechanisms that govern the optical response. In this work, we demonstrate the adaptation of the transmission line model approach for VdWHs, based on expressing the VdWH constituents by distributed electrical circuit elements described by their admittance. Since the admittance carries fundamental physical meaning of the material response to electromagnetic fields, the approach results in a one-dimensional system of propagating voltage and current waves, offering a compact and physically intuitive formulation that simplifies algebraic calculations, clarifies the conditions for existence of physical solutions, and provides valuable insight into the fundamental physical response. To demonstrate the robustness and advantages of this approach, we derive the transmission line analogs of bulk to monolayer 2D materials and show how the approach can be used to compute the reflection/transmission coefficients, polaritonic dispersion relations, and electromagnetic field distributions in a variety of 2D material VdWHs, and compare them to experimental measurements yielding very good agreement. This method provides a valuable tool for exploring and understanding the optical response of layered 2D systems.
\end{abstract}

\maketitle

\section{Introduction}

Van der Waals heterostructures (VdWHs) are structures composed of stacked layers of different two-dimensional (2D) materials, each potentially differing in thickness and composition \cite{Geim2013VanHeterostructures, Castellanos-Gomez2022VanHeterostructures}. These 2D materials span a broad range of material type classification, such as semi-metallic (graphene) \cite{Geim2007TheGraphene, Grigorenko2012, Geim2009Graphene:Prospects}, dielectric (hexagonal boron nitride (hBN)) \cite{Caldwell2019PhotonicsNitride, Zhang2017TwoApplications, Zunger1976OpticalNitride}, semiconducting (transition metal dichalcogenides (TMDs)) \cite{Manzeli20172DDichalcogenides, Wang2012ElectronicsDichalcogenides, Choi2017RecentApplications} and more. 
The combination of this compositional diversity, intrinsic low dimensionality, strong quantum confinement effects, inherent anisotropy and integrability of individual 2D materials into VdWH results in unique and tunable optical response \cite{Xia2014Two-dimensionalNanophotonics, Grigorenko2012, Wang2018Colloquium:Dichalcogenides, Caldwell2015Low-lossPolaritons, Caldwell2019PhotonicsNitride}.

The interaction between light and 2D materials and their heterostructures has been extensively investigated, experimentally and theoretically \cite{Geim2013VanHeterostructures, Deilmann2020LightmatterHetero-structures, Schneider2018Two-dimensionalCoupling, Deilmann2020LightmatterHetero-structures, Koppens2014PhotodetectorsSystems, Akinwande2019GrapheneTechnology, Ferrari2015, deAbajo2025RoadmapMaterials}.
A central aspect of this interaction is the response of the heterostructure to impinging electromagnetic waves, resulting in quasiparticle excitations in the material and a unique reflection/transmission response. Of particular interest in 2D materials are polaritons — quasiparticles that arise from the interaction between electromagnetic fields and material excitations — combining characteristics of both the photon and the material excitation \cite{Basov2016PolaritonsMaterials, Low2016, Basov2021PolaritonPanorama}, that can be studied through the analysis of electromagnetic eigenmodes in the system.

To study these type of optical and polaritonic properties of VdWHs, different methods have been applied via the analysis of Maxwell’s equations, such as the transfer matrix method (TMM) that relates the input and output electromagnetic fields via transmission and propagation matrices \cite{Zhan2013TransferLayers, BornPrinciplesLight}, or the multiple reflection summation method, where the total reflection and transmission coefficients are expressed as an infinite series of round-trip contributions involving the local Fresnel coefficients at each interface \cite{BornPrinciplesLight, ChoChew1995WavesMedia}, and related techniques. While these methods are straight forward and have been commonly used in the study of light interaction with 2D material VdWHs, they often yield long and complicated analytical expressions, especially in multilayer heterostructures, which can obscure the underlying physical mechanisms of the response, giving limited intuitive insight into the light–matter interaction in the heterostructure.

In this work, we propose an improved, physically informed description of the optical response of VdWHs using the framework of a transmission line model (TLM) \cite{DavidkChengField-and-wave-electromagnetics, PozarD.M.2011MicrowaveEngineering, Schelkunoff1937TransmissionWaves}. In this approach, the VdWH constituents are described by distributed electrical circuit elements, characterized by their admittance. This approach is highly advantageous as the admittance carries fundamental physical meaning of the material response to electromagnetic fields. Together with the fact that it results in a one-dimensional system of propagating voltage and current waves, the TLM yields a compact and physically intuitive formulation that simplifies algebraic calculations, clarifies the conditions for existence of physical solutions, and provides valuable insight into the governing physics of the response. 
The TLM is a useful method due to the ability to relate between material properties and transmission line characteristics \cite{Marcuvitz1951WaveguideHandbook}. 
For example, \cite{Lee2012IntrinsicFrequencies} used the TLM to analyze the characteristics of graphene. While the TLM has been used previously in 2D material structures, the potential of this method has not been realized yet in the 2D community.
To demonstrate the robustness of the TLM approach, we derive the transmission line analogs of bulk to monolayer 2D materials and their heterostructures, and compute the reflection and transmission coefficients, polaritonic dispersion relations, and electromagnetic field distributions in various VdWHs. These VdWHs include hBN-encapsulated TMD, demonstrating the effect of TMD excitons on reflection; hBN-encapsulated biased graphene, with strongly confined polaritons; and hBN encapsulated with a dielectric material, supporting hyperbolic polaritons. In addition, we show that the optical response predicted by the TLM agrees well with experiential measurements.

\section{Modeling VdWHs with Transmission Lines}
\label{TLM}
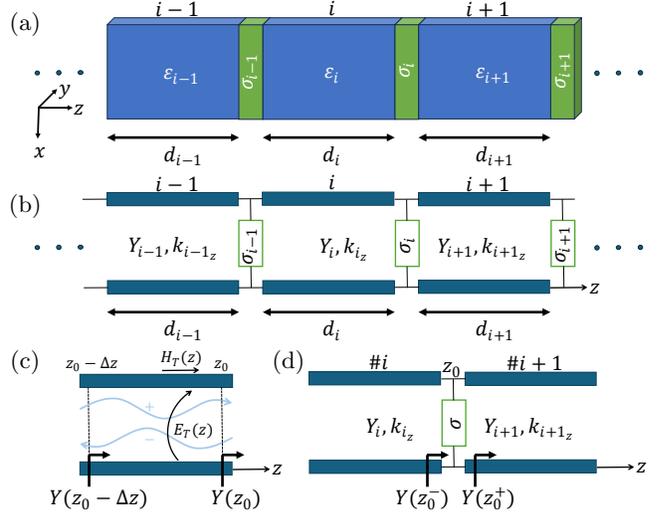
\begin{figure}[t]
    \centering
    \hspace*{-13pt} 
    \begin{tikzpicture}
    \node[anchor=south west,inner sep=0] (image) at (0,0) {\includegraphics
    [width=0.5\textwidth]
    {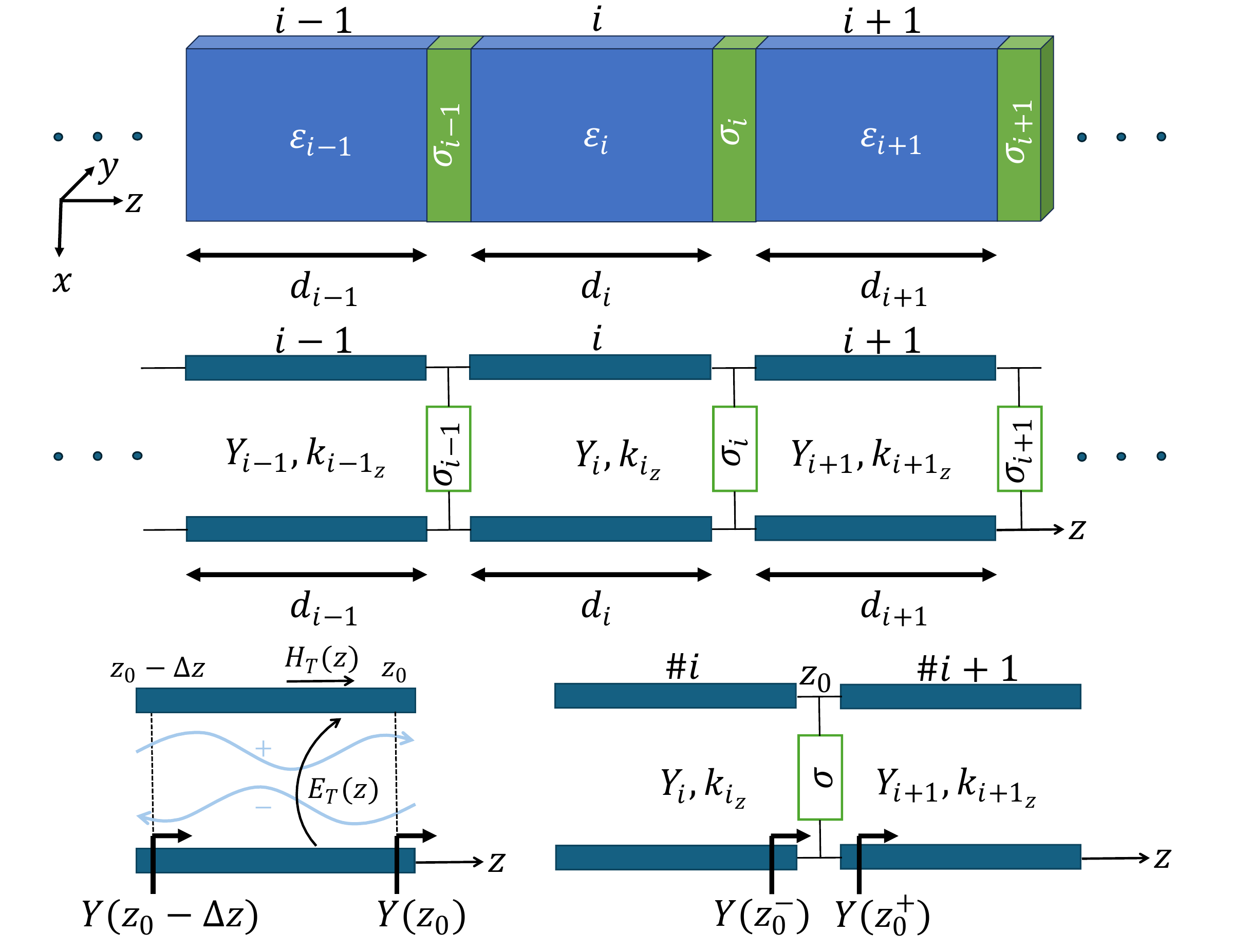}};
    \begin{scope}[x={(image.south east)},y={(image.north west)}]
    \node at (0.03,0.95) {(a)};
    \node at (0.03,0.6) {(b)};
    \node at (0.03,0.3) {(c)};
    \node at (0.42,0.3) {(d)};
    \end{scope}
    \end{tikzpicture}
    \caption{Transmission line model for VdWHs. (a) A VdWH configuration constructed by layers of bulk, few-layer, or monolayer 2D materials. (b) Transmission line model of the VdWH in (a), where bulk layers are modeled as a transmission line and thin layers as a parallel admittance. (c) Transmission line model of the $i$th layer showing the transverse fields as analog to the voltage and current, propagating through forward and backward waves. The $z$ dependent admittance at two different locations in the layer are marked. (d) A thin layer in the VdWH modeled as a parallel admittance between transmission lines.}
    \label{fig def}
\end{figure}

We assume a layered 2D material heterostructure constructed of uniaxial anisotropic layers (Fig. \ref{fig def} (a)), such that the permittivity of the $i$th bulk layer is given by $\varepsilon_{i}=diag(\varepsilon_{i_{xx}},\varepsilon_{i_{xx}},\varepsilon_{i_{zz}})$, where $\varepsilon_{i_{xx}}$ and $\varepsilon_{i_{zz}}$ are the in-plane and out-of-plane permittivity components, respectively. Due to a symmetry in the $x-y$ plane, assuming the VdWH is infinite in that plane, we take the propagation direction to be $\hat{x}$ and assume uniformity of the fields in the $\hat{y}$ direction without loss of generality. A field time dependence of $e^{-i\omega t}$ is assumed as well, with a transverse dependence of $e^{iqx}$, where $q$ is the $x$ component of the wavevector (momentum), allowed due to the continuity of the transverse electric field on the boundaries between materials (see section \ref{BC thin layer}). The resulting fields are divided into the TM fields: 
$\mathbb{E}=(\hat{x}\mathbb{E}_x(z)+\hat{z}\mathbb{E}_z(z))e^{iqx}$ and $\mathbb{H}=\hat{y}\mathbb{H}_y(z)e^{iqx}$, and the TE fields: 
$\mathbb{H}=(\hat{x}\mathbb{H}_x(z)+\hat{z}\mathbb{H}_z(z))e^{iqx}$ and $\mathbb{E}=\hat{y}\mathbb{E}_y(z)e^{iqx}$.

Using Maxwell's equations, the above transverse electric $\mathbb{E}_T(z)$ and magnetic $\mathbb{H}_T(z)$ fields ($\mathbb{E}_x(z)$, $\mathbb{H}_y(z)$ for TM and $\mathbb{E}_y(z)$, $-\mathbb{H}_x(z)$ for TE), satisfy the telegraph equations describing voltage $V(z)$ and current $I(z)$ in a transmission line \cite{DavidkChengField-and-wave-electromagnetics, PozarD.M.2011MicrowaveEngineering, Schelkunoff1937TransmissionWaves, Felsen1994RadiationWaves}:
\begin{align}
    \frac{d\mathbb{E}_T(z)}{dz}=ik_{i_z}Y_i^{-1} \mathbb{H}_T(z) ,
\\
    \frac{d\mathbb{H}_T(z)}{dz}=ik_{i_z}Y_i \mathbb{E}_T(z) ,
\end{align}
where $k_{i_z}$ is the $z$ component of the wavevector (propagation coefficient) in the $i$th layer, related to the momentum by \cite{Ferrari2015HyperbolicApplications} $\frac{q^2}{\varepsilon_{i_{zz}}}+\frac{k_{i_z}^2}{\varepsilon_{i_{xx}}}=k_0^2$ and $q^2+k_{i_z}^2=\varepsilon_{i_{xx}}k_0^2$ for TM and TE, respectively, where $k_0$ is the free-space wavenumber, and $Y_i$ is the characteristic admittance of the $i$th layer, equals to $\frac{\omega \varepsilon_0 \varepsilon_{i_{xx}}}{k_{i_z}}$ and $\frac{k_{i_z}}{\omega \mu_0}$ for TM and TE, respectively. Therefore, each bulk 2D material in the VdWH can be modeled as a transmission line (Fig \ref{fig def} (b)), and the transverse electric and magnetic fields will be referred to as voltage and current in further discussions. 

The transmission line supports forward ($+$) and backward ($-$) propagating waves, with a longitudinal dependence of $e^{\pm i k_{i_z} z}$, with characteristic admittance of a transmission line relating the forward or backward propagating current and voltage via
\begin{align}
    Y_i=\frac{\mathbb{H}^+_T(z)}{\mathbb{E}^+_T(z)}=-\frac{\mathbb{H}^-_T(z)}{\mathbb{E}^-_T(z)} .
\end{align}
The $z$ dependent admittance is defined as the ratio between the current and the voltage: $Y(z)=\frac{\mathbb{H}_T(z)}{\mathbb{E}_T(z)}$. 
If the $i$th layer is half infinite for $z \rightarrow \pm \infty$, and there is no external impinging wave in the layer, then there is only a forward (or backward) propagating wave, and in this layer: $Y(z)=\pm Y_i$.
The reflection coefficient at $z$ in the $i$th layer is defined as:
\begin{align}
    \label{Eq ref coef}
    r_i(z)=\frac{\mathbb{E}^-_T(z)}{\mathbb{E}^+_T(z)}=-\frac{\mathbb{H}^-_T(z)}{\mathbb{H}^+_T(z)} ,
\end{align}
is given by:
\begin{align}
\label{Eq ref}
    r_i(z) = \frac{Y_i-Y(z)}{Y_i+Y(z)} .
\end{align}

The $z$ dependent admittance at two different locations in the same $i$th layer is related by a tangent equation (Fig. \ref{fig def} (c)). Using the $z$ dependent admittance at the $i$th layer given by:
\begin{align}
    \label{Eq admittance using current}
    Y(z)=\frac{\mathbb{H}^+_T(z)+\mathbb{H}^-_T(z)}{\mathbb{E}^+_T(z)+\mathbb{E}^-_T(z)}=Y_{i}\frac{\mathbb{H}^+_T(z)+\mathbb{H}^-_T(z)}{\mathbb{H}^+_T(z)-\mathbb{H}^-_T(z)} ,
\end{align}
and inserting the reflection coefficient (Eq. \ref{Eq ref coef} and \ref{Eq ref}) and the $z$ dependence of the waves into Eq. \ref{Eq admittance using current} at $z - \Delta z$ gives the tangent equation:
\begin{align}
    \label{Eq tangent}
    \begin{split}
    Y(z - \Delta z) &=Y_{i} \frac{\mathbb{H}^+_T(z) e^{-i k_{i_z} \Delta z}+\mathbb{H}^-_T(z) e^{i k_{i_z} \Delta z}}{\mathbb{H}^+_T(z) e^{-i k_{i_z} \Delta z}-\mathbb{H}^-_T(z) e^{i k_{i_z} \Delta z}} = \\
    & = 
    Y_{i} \frac{Y(z) - i Y_{i} \tan(k_{i_z} \Delta z)}{Y_{i} - i Y(z) \tan(k_{i_z} \Delta z)} .
    \end{split}
\end{align}

\section{Boundary conditions of a thin layer}
\label{BC thin layer}
Assuming a thin layer (usually a monolayer) described by its conductivity $\sigma$ located at $z_0$ in the VdWH, and modeled as an infinitesimal layer with surface conductivity $\sigma$ \cite{Hanson2008DyadicGraphene}, the boundary conditions at $z_0$ are given by: $\mathbb{E}_T(z_0^-)=\mathbb{E}_T(z_0^+)$ and $\mathbb{H}_T(z_0^-)-\mathbb{H}_T(z_0^+)=\sigma \mathbb{E}_T(z_0)$. Note that these boundary conditions also agree with the transverse dependence introduced in section \ref{TLM}, and in terms of admittances are given by:
\begin{equation}
Y(z_0^-)-Y(z_0^+)=\sigma ,
\label{eq admittance BC}
\end{equation}
the thin layer can thus be modeled as a parallel admittance between the transmission lines, as described in Fig. \ref{fig def} (d).

We note that the conductivity, under the assumption of a thin layer, is related to the in-plane susceptibility of the thin layer $\chi_{xx}$
through $\sigma=-i\omega \varepsilon_0 \chi_{xx}d$, where $\omega=\frac{E}{\hbar}$ is the frequency and $d$ is the thickness of the thin layer \cite{Senior1995ApproximateElectromagnetics}. 

\section{Reflection from a VdWH}

\begin{figure}[t]
    \centering
    \hspace*{-13pt} 
    \begin{tikzpicture}
    \node[anchor=south west,inner sep=0] (image) at (0,0) {\includegraphics[width=0.5\textwidth]{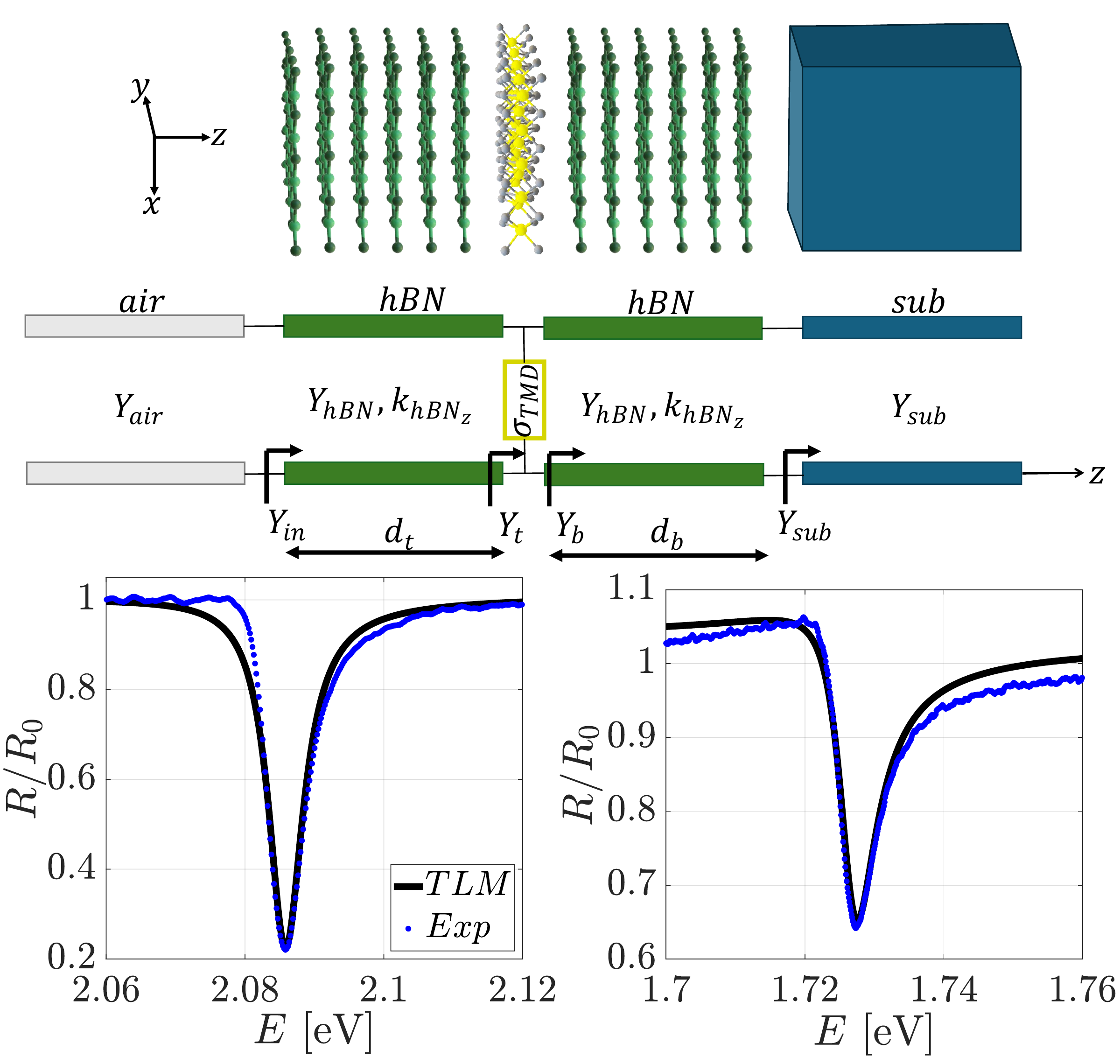}};
    \begin{scope}[x={(image.south east)},y={(image.north west)}]
    \node at (0.03,0.94) {(a)};
    \node at (0.03,0.75) {(b)};
    \node at (0.03,0.43) {(c)};
    \node at (0.52,0.43) {(d)};
    \end{scope}
    \end{tikzpicture}
    \caption{Transmission line model for excitonic reflection from a $\mathrm{WS_2}$ and $\mathrm{WSe_2}$ VdWHs. (a) An illustration of the hBN/monolayer TMD/hBN/substrate VdWH. (b) A transmission line scheme for the configuration in (a). Measured reflection contrast (blue points) and TLM fit black lines) for (c) the $\mathrm{WS_2}$ VdWH on a gold mirror at $T=4K$, and(d) $\mathrm{WSe_2}$ VdWH on $\mathrm{SiO_2}$/Si substrate at $T=80K$. }
    \label{fig ref}
\end{figure}
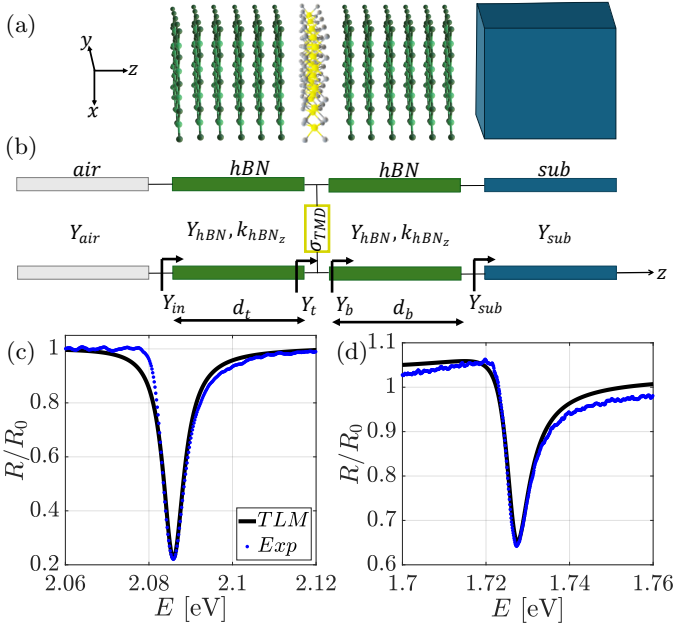

As an exemplary case for calculating the reflection from a VdWH using the TLM, we examine the hBN/monolayer TMD/hBN/substrate heterostructure commonly used for investigating the optical response of excitons in monolayer TMD heterostructures \cite{Kahlon2025ImportanceHeterostructures, Epstein2020Near-UnityCavity, Epstein2020, Horng2020PerfectCrystal, Robert2018OpticalHeterostructures, Rogers2020CoherentExcitons, Li2021RefractiveDiselenide, Goncalves2018Plasmon-excitonInterfaces, Ajayi2017ApproachingMonolayers, Cadiz2017ExcitonicHeterostructures} 
(Fig. \ref{fig ref} (a)). Since perpendicular incidence with TM polarization was assumed, we have $k_{i_z}=\sqrt{\varepsilon_{i_{xx}}}k_0$ and $Y_i= c \varepsilon_0\sqrt{\varepsilon_{i_{xx}}}$, where $c$ is the speed of light in vacuum. The $z$ component of the wavevector in the hBN is denoted as $k_{hBN_z}$ and the characteristic admittances of air, hBN and the substrate are denoted as $Y_{air}$, $Y_{hBN}$ and $Y_{sub}$, respectively. Figure \ref{fig ref} (a) and (b) present the VdWH composition and TLM model, respectively. The  admittance at the interface between the substrate and bottom (right) hBN layers is given by $Y_{sub}$, since the substrate is modeled as half infinite in the region $z>0$. The admittance at the interface between the TMD and bottom hBN is obtained using Eq. \ref{Eq tangent}
\begin{align}
    Y_{b}=Y_{hBN} \frac{Y_{sub} - i Y_{hBN} \tan(k_{hBN_z} d_{b})}{Y_{hBN} - i Y_{sub} \tan(k_{hBN_z} d_{b})} ,
\end{align}
where $d_b$ is the thickness of the bottom hBN. The admittance at the interface between the top (left) hBN and the TMD is given by $Y_{t}=Y_{b}+\sigma_{TMD}$, where $\sigma_{TMD}$ is the conductivity of the TMD monolayer. The admittance at the interface between the air and top hBN (the entrance admittance) is given by:
\begin{align}
    Y_{in}=Y_{hBN} \frac{Y_{t} - i Y_{hBN}\tan(k_{hBN_z} d_{t})}{Y_{hBN} - i Y_{t} \tan(k_{hBN_z} d_{t})} ,
\end{align}
where $d_t$ is the thickness of the top hBN. The total reflection coefficient is therefore $r = \frac{Y_{air}-Y_{in}}{Y_{air}+Y_{in}}$.

To validate our theoretical model, we have measured the reflection coefficient from the two different VdWHs, in the first the TMD is $\mathrm{WS_2}$ at $T=4K$ above mirror gold substrate (Fig. \ref{fig ref} (c)) and in the second the TMD is $\mathrm{WSe_2}$ at $T=80K$ above $\mathrm{SiO_2}$/Si substrate (Fig. \ref{fig ref} (d)). We then calculate the reflection contrast, $R/R_0$, where $R=|r|^2$, $R_0=|r_0|^2$; $r$ and $r_0$ are the reflection coefficients with and without the presence of the TMD, respectively. The air above the VdWH is modeled as an additional transmission line since the top hBN has a finite thickness and the conductivity of the monolayer TMD is calculated through its in-plane susceptibility \cite{Li2014MeasurementDichalcogenides, Scuri2018LargeNitride, Back2018RealizationMoSe2, Shahmoon2017CooperativeArrays,Horng2020PerfectCrystal, Robert2018OpticalHeterostructures}:
\begin{align}
    \chi_{xx}=\chi_{bg}-\frac{c}{\omega_0d_0}\frac{\gamma_{r}}{\omega-\omega_0+i\frac{\gamma_{nr}}{2}} ,
\end{align}
where $\chi_{bg}$ is the background susceptibility, $\omega_0$ is the exciton resonance frequency $d_0$ is the monolayer thickness, $\gamma_{r}$ and  $\gamma_{nr}$ are the radiative and non-radiative decay rates, respectively. The susceptibility parameters are fitted to the measurements as described in \cite{Kahlon2025ImportanceHeterostructures, Epstein2020Near-UnityCavity}.
The contribution of the silicon in the second sample is neglected, since the parameters chosen in the experiment satisfy $|Im(k_{{SiO_2}_z})d_{SiO_2}|\gg1$, the signal significantly decays until it reaches the $\mathrm{SiO_2}$/Si interface, and therefore no significant reflection from the interface exists. Figure \ref{fig ref} (c) and (d) present the TLM calculated (black line) and the measured (blue points) reflection contrast for the two samples, showing good agreement.

\section{Transverse Resonance Equation}
\label{section TRE} 

The dispersion relation associated with the electromagnetic eigenmodes propagating in the VdWH, and in particular polaritons, can be found by solving the transverse resonance equation, obtained by equalizing the solutions of the $z$ dependent admittance at a point $z_0$ that satisfy the boundary conditions for $z>z_0$ and $z<z_0$, when no external wave is present \cite{PozarD.M.2011MicrowaveEngineering}. If the point $z_0$ is taken as the interface between the first and second layers in Fig. \ref{fig def} (a), then the transverse resonance equation is $Y_{tot}=- Y_1$, where $Y_{tot}$ satisfies the boundary condition for $z>z_0$. From Eq. \ref{Eq ref} it is clear that the transverse resonance equation is equivalent for taking the total reflection coefficient to infinity $r_{tot}=\frac{Y_1-Y_{tot}}{Y_1+Y_{tot}} \rightarrow \infty$, as the poles of the total reflection coefficient are associated with the eigen modes of the heterostructure \cite{Alpeggiani2017Quasinormal-modeMatrix, Leung1994CompletenessCavities, Muljarov2011Brillouin-WignerSystems, Kristensen2012GeneralizedCavities, Sauvan2013TheoryResonators, Muljarov2016Resonant-stateSand}.

We can further simplify this formulation since polaritons are characterized by high in-plane momentum \cite{Basov2016PolaritonsMaterials, Low2016, Basov2021PolaritonPanorama}. 
This allows to approximate $q \gg \sqrt{\varepsilon_{i_{zz}}}k_0$, leading to $k_{i_z}\approx i q \sqrt{\frac{\varepsilon_{i_{xx}}}{\varepsilon_{i_{zz}}}}$ for TM polaritons. Therefore, the characteristic admittance can be written as $Y_i=\frac{\omega \varepsilon_0 \varepsilon_{i_{eff}}}{i q }$, where $\varepsilon_{i_{eff}}=\sqrt{\varepsilon_{i_{xx}}\varepsilon_{i_{zz}}}$ is the effective permittivity of the $i$th bulk layer. A condition for the existence of polaritons in a heterostructure is that at least one of the layers has a negative real part of the permittivity (or a positive imaginary part of the conductivity) \cite{MaierPLASMONICS:APPLICATIONS, Epstein2020, Eini2022, Eini2025ElectricallyGrapheneb, Eini2025Exciton--hyperbolic-phonon-polaritonGraphene, Sternbach2021ProgrammableSemiconductors, Li2014}. 

\section{Surface polaritons in a VdWH}
\label{section GEP}
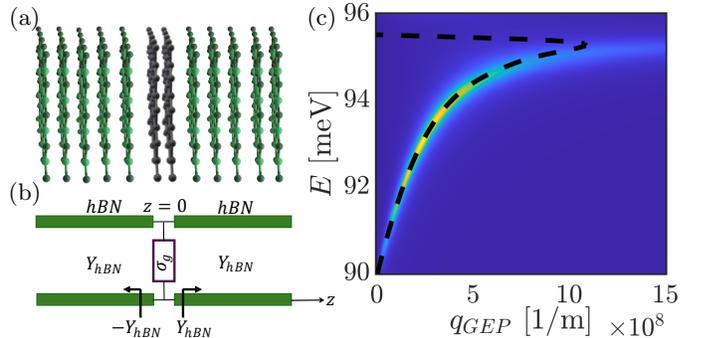
\begin{figure}[b]
    \centering
    \hspace*{-13pt} 
    \begin{tikzpicture}
    \node[anchor=south west,inner sep=0] (image) at (0,0) {\includegraphics[width=0.5\textwidth]{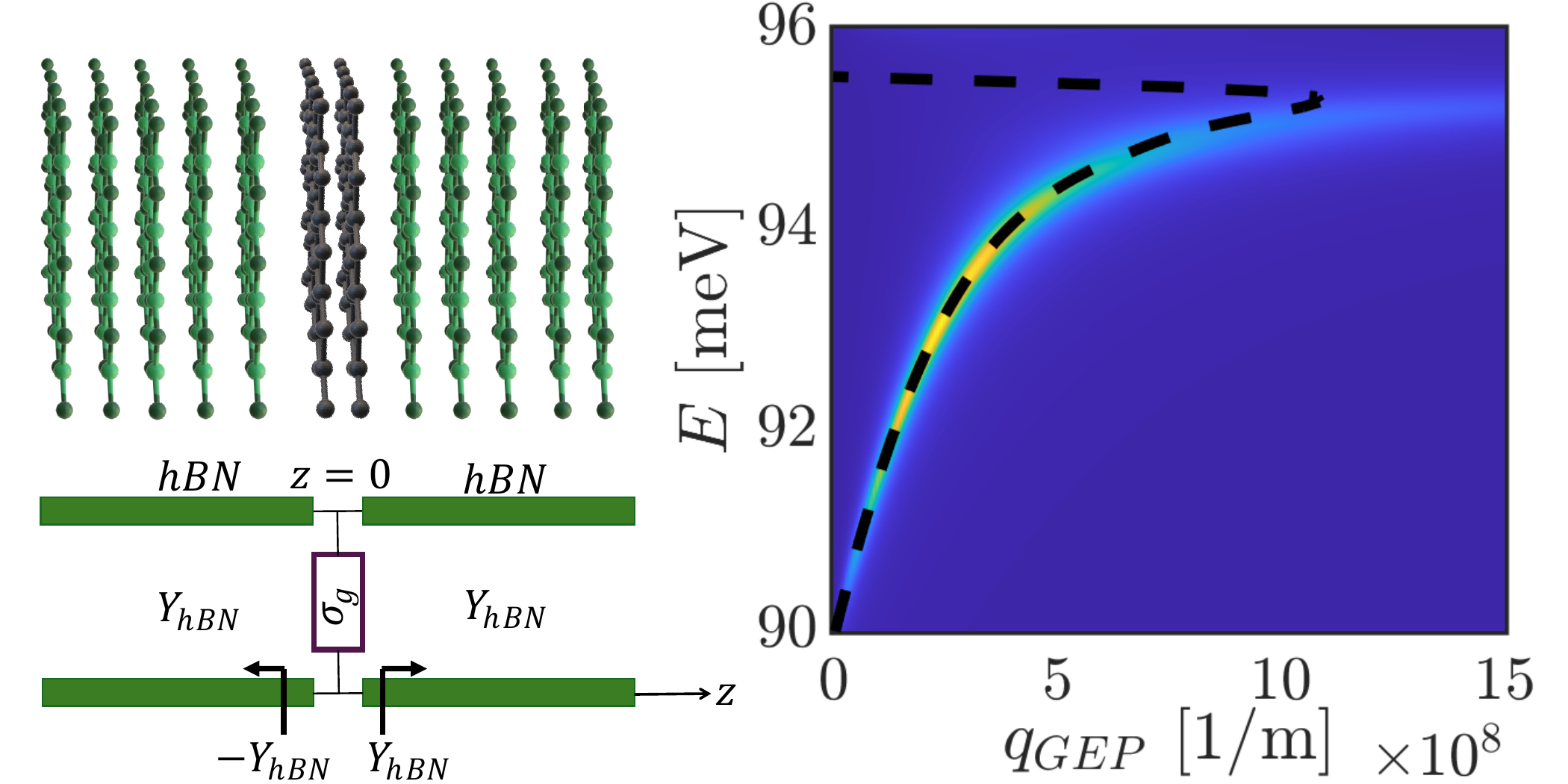}};
    \begin{scope}[x={(image.south east)},y={(image.north west)}]
    \node at (0.01,0.95) {(a)};
    \node at (0.01,0.43) {(b)};
    \node at (0.45,0.95) {(c)};
    \end{scope}
    \end{tikzpicture}
    \caption{Transmission line model for GEP. (a) An illustration of the hBN/BLG/hBN VdWH. (b) Transmission line schemes for the configuration. (c) Dispersion relation of GEP in the configuration, calculated from Eq. \ref{Eq GEP DR} (dashed black line) compared to TMM simulation (colormap).}
    \label{fig GEP}
\end{figure}

As an exemplary case of obtaining a polaritonic dispersion relation, we derive the dispersion relation of graphene-exciton-polaritons (GEPs) supported by a hBN/bilayer graphene (BLG)/hBN heterostructure \cite{Eini2025ElectricallyGrapheneb} (Fig. \ref{fig GEP} (a)). The $z$ dependent admittance at $z \neq 0$, where the BLG is located at $z=0$, is given by $Y(z)=sign(z) Y_{hBN}$, since the hBN layers are half infinite for $z \rightarrow \pm \infty$. The BLG is modeled as an infinitesimal layer with conductivity $\sigma_g$, therefore $Y(0^-)-Y(0^+)=\sigma_g$, resulting in the transverse resonance equation: 
\begin{equation}
    2Y_{hBN}+\sigma_g =0 ,
    \label{Eq GEP admittace matching}
\end{equation}
where $Y_{hBN}=\frac{\omega \varepsilon_0 \varepsilon_{hBN_{eff}}}{i q }$. Assuming $\varepsilon_{hBN_{eff}}$ is real and positive, which holds outside its Reststrahlen bands \cite{Dai2014TunableNitride}, it can be deduced that $Y_{hBN}$ is capacitive in nature (negative imaginary admittance). Hence $\sigma_g$ must be inductive (positive values of its imaginary part) in order to satisfy Eq. \ref{Eq GEP admittace matching}, which is enabled due to the excitonic response of the biased BLG systems \cite{Park2010TunableGraphene, Cao2018UnifyingNumbers, Ju2017TunableGraphene,Ju2020UnconventionalGraphene, Henriques2022AbsorptionGraphene, Quintela2022TunableGraphene, Eini2025ElectricallyGrapheneb}. Inserting the expression of $Y_{hBN}$ into Eq. \ref{Eq GEP admittace matching} gives the known equation for a TM surface polariton in a 2D infinitesimal layer
\begin{equation}
    q_{GEP} =\frac{2i\omega \varepsilon_0 \varepsilon_{hBN_{eff}}}{\sigma_g } ,
    \label{Eq GEP DR}
\end{equation}
as have been previously obtained for graphene plasmons in monolayer graphene \cite{Goncalves2016AnPlasmonics} and for exciton polaritons in monolayer TMD \cite{Epstein2020HighlySemiconductors, Eini2022, Kats2025MonolayerFrequencies, Eini2025ElectricallyGrapheneb},
for example, using other methods. Figure \ref{fig GEP} (a) and (b) present the VdWH configuration and its TLM model, respectively. Figure \ref{fig GEP} (c) compares the dispersion relation of the GEP in the VdWH calculated using Eq. \ref{Eq GEP DR} (dashed black line) and simulated using TMM (color map) showing excellent agreement.

\section{Hyperbolic polaritons in a VdWH}
\label{section HPhP}
\begin{figure}[t]
    \centering
    \hspace*{-13pt} 
    \begin{tikzpicture}
    \node[anchor=south west,inner sep=0] (image) at (0,0) {\includegraphics[width=0.5\textwidth]{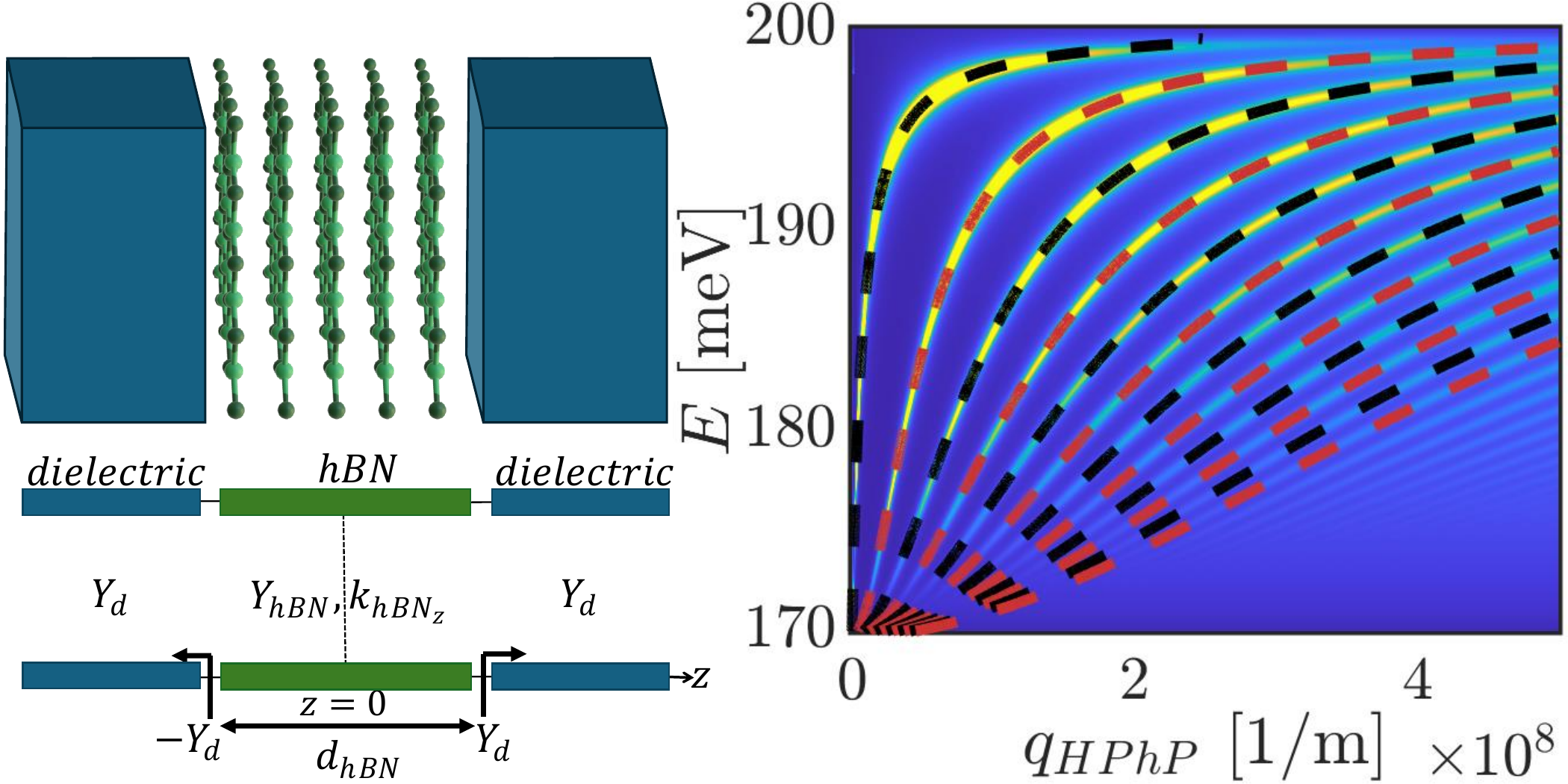}};
    \begin{scope}[x={(image.south east)},y={(image.north west)}]
    \node at (0.00,0.97) {(a)};
    \node at (0.00,0.43) {(b)};
    \node at (0.44,0.97) {(c)};
    \end{scope}
    \end{tikzpicture}
    \caption{Transmission line model for HPhP. (a) An illustration of the dielectric/hBN/dielectric VdWH. (b) Transmission line schemes for the configuration in (a). (c) Dispersion relation of HPhP in the configuration, calculated from Eq. \ref{Eq HPhP DR} (dashed black line for even modes and dashed red line for odd modes) compared to TMM simulation (colormap).}
    \label{fig HPhP}
\end{figure}
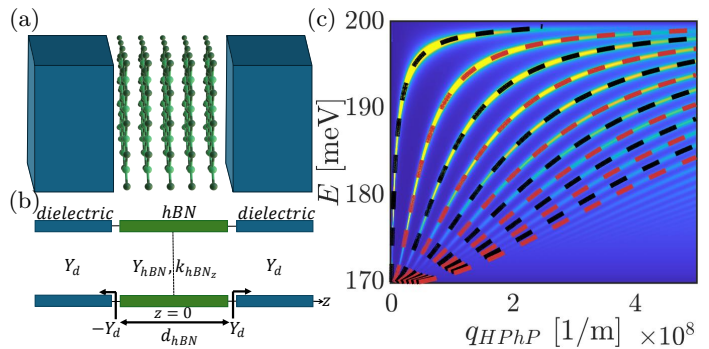

Next we explore the response of propagating Hyperbolic-Phonon-Polaritons (HPhPs) in hBN, which have been of a large scientific interest in recent years owing to their ability to exhibit large confinement factors and low propagation losses \cite{Caldwell2014Sub-diffractionalNitride, Dai2014TunableNitride, Li2015HyperbolicFocusing, Zheng2018HighlyOxides}. 
We derive the dispersion relation of HPhPs supported in a dielectric/hBN/dielectric heterostructure (Fig. \ref{fig HPhP} (a)) using the equivalent TLM presented in Fig. \ref{fig HPhP} (b), while previous approaches have used Fresnel equations and ray-based methods \cite{Dai2014TunableNitride, Zheng2018HighlyOxides}, for exmample. 
The $z$ dependent admittance at $|z|>\frac{d_{hBN}}{2}$, where the hBN is located at $|z|<\frac{d_{hBN}}{2}$, is given by $Y(z)=sign(z) Y_{d}$, where $Y_{d}$ is the characteristic admittance of the dielectric, since the dielectric layers are half infinite for $z \rightarrow \pm \infty$. Using Eq. \ref{Eq tangent} to relate between $Y(\frac{d_{hBN}}{2})$ and $Y(-\frac{d_{hBN}}{2})$ we get:
\begin{align}
    -Y_{d} =
    Y_{hBN} \frac{Y_{d} - i Y_{hBN} \tan(k_{hBN_z} d_{hBN})}{Y_{hBN} - i Y_{d} \tan(k_{hBN_z} d_{hBN})} .
    \label{eq:qHphp}
\end{align}
By setting $Y_{hBN}=\frac{\omega \varepsilon_0 \varepsilon_{hBN_{eff}}}{i q }$, $Y_{d}=\frac{\omega \varepsilon_0 \varepsilon_{d_{eff}} }{i q }$ and $k_{hBN_z}\approx i q \sqrt{\frac{\varepsilon_{hBN_{xx}}}{\varepsilon_{hBN_{zz}}}}$, after some algebra we obtain the known dispersion relation of hyperbolic polaritons, as previously  obtained for HPhPs and Hyperbolic-Exciton-polaritons using other methods \cite{Dai2014TunableNitride, Eini2022} : 
\begin{align}
\label{Eq HPhP DR}
    q_{HPhP}=\frac{i}{d_{hBN}}\sqrt{\frac{\varepsilon_{hBN_{zz}}}{\varepsilon_{hBN_{xx}}}}\bigg[2\arctan \bigg(i\frac{\varepsilon_{d_{eff}}}{\varepsilon_{hBN_{eff}}} \bigg)+\pi L \bigg] ,
\end{align}
where $L$ is the modal order.

Due to the symmetry of the VdWH around the $z$ axis, it supports even modes (even voltage around the $z$ axis) and odd modes (odd voltage around the $z$ axis). We note that the parity of the current is opposite to the parity of the voltage in TM polarization. We can use this symmetry to obtain more easily the dispersion relation in Eq. \ref{Eq HPhP DR}.

Since the transverse fields are continuous at $z=0$, the even modes satisfy a nullification of the current at $z=0$ and $Y(z=0)=0$ and the odd modes satisfy a nullification of the voltage at $z=0$ and $Y^{-1}(z=0)=0$. Applying these conditions to Eq. \ref{eq:qHphp}, we obtain for the even modes $Y_{air} - i Y_{hBN} \tan(k_{even,hBN_z} \frac{d_{hBN}}{2})=0$ and for the odd modes is given by $Y_{hBN} - i Y_{air} \tan(k_{odd,hBN_z} \frac{d_{hBN}}{2})=0$. From these equations, the dispersion relation in Eq. \ref{Eq HPhP DR} can be obtained, with even modal orders for the even modes and odd modal orders for the odd modes.

Figure \ref{fig HPhP} (c) compares the dispersion relation of the HPhP in the VdWH calculated using Eq. \ref{Eq HPhP DR} (dashed black line for even modes and dashed red line for odd modes) and simulated using the TMM (color map), showing very well agreement.

As another robust example of using the TLM to analyze the hyperbolic response of a VdWh, we note that Kats et. al. \cite{Kats2025MonolayerFrequencies} have recently shown that an alternating TMD monolayers and hBN layered VdWH exhibits a hyperbolic response, by analyzing the reflection and polaritonic properties of the structure as obtained from TLM approach.

\section{Polaritons Hybridization in a VdWH}

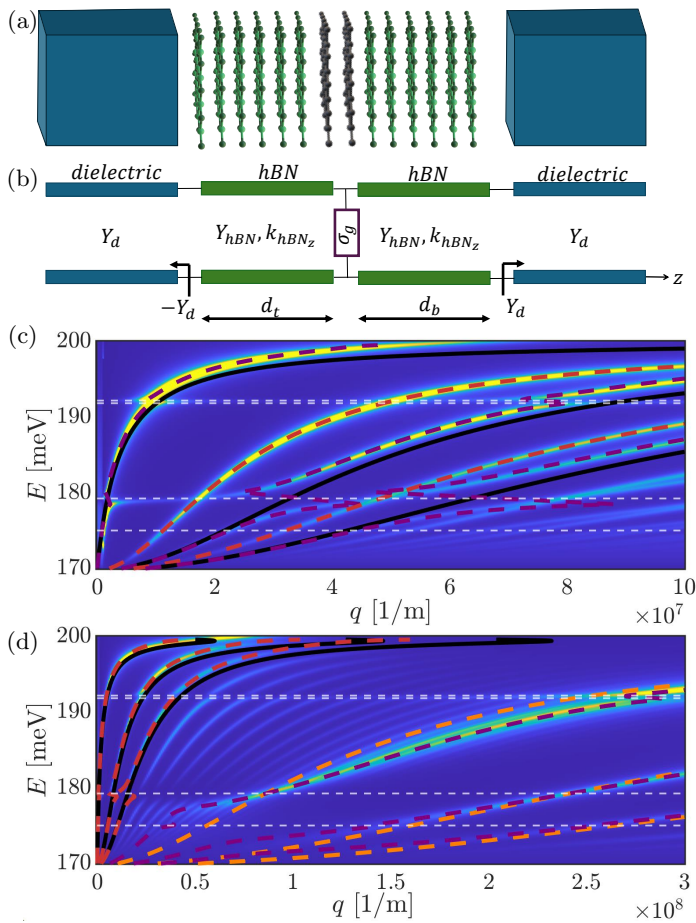
\begin{figure}[b]
    \centering
    \hspace*{-13pt} 
    \begin{tikzpicture}
    \node[anchor=south west,inner sep=0] (image) at (0,0) {\includegraphics[width=0.5\textwidth]{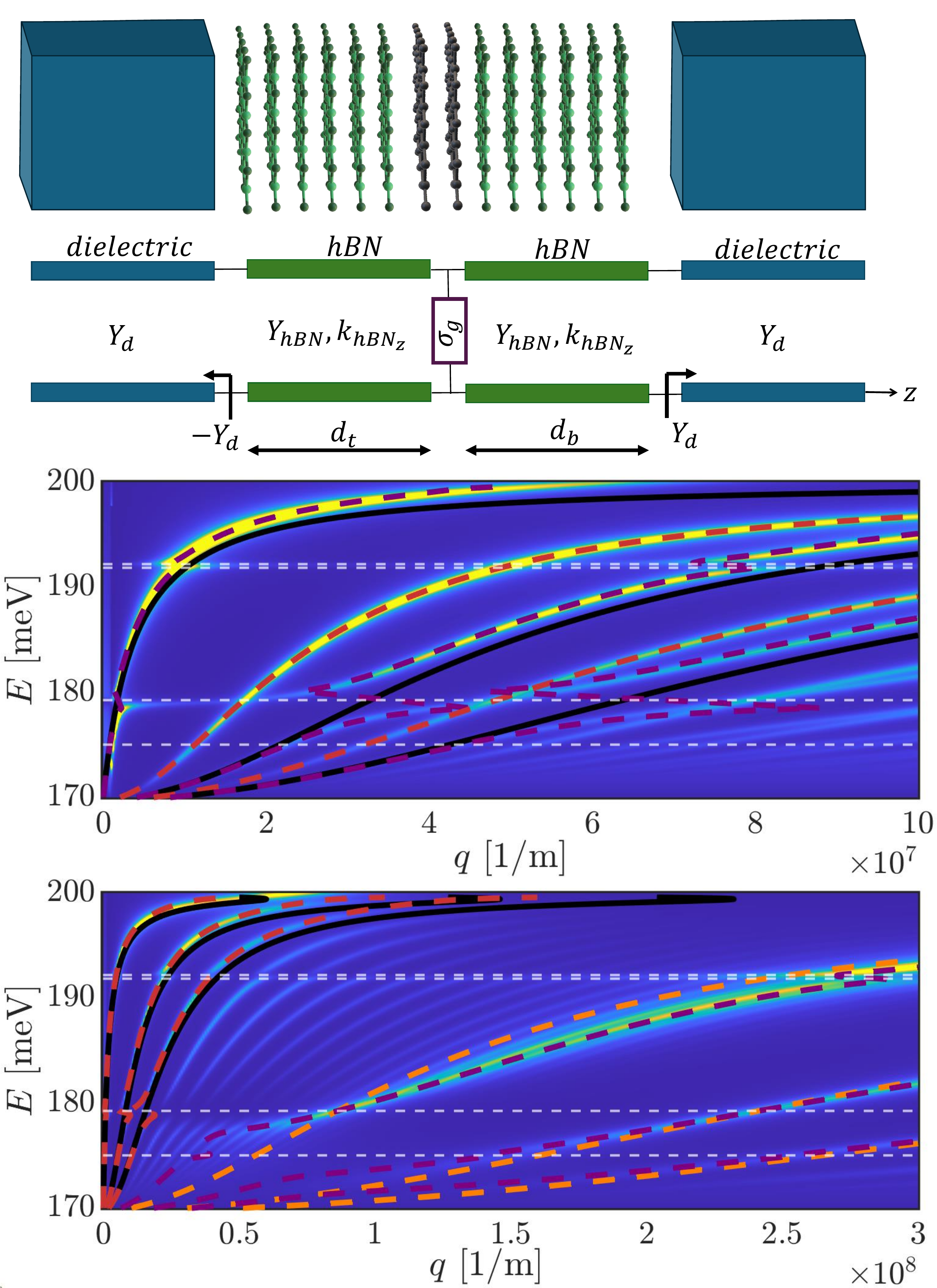}};
    \begin{scope}[x={(image.south east)},y={(image.north west)}]
    \node at (0,0.97) {(a)};
    \node at (0,0.8) {(b)};
    \node at (0,0.63) {(c)};
    \node at (0,0.3) {(d)};
    \end{scope}
    \end{tikzpicture}
    \caption{Transmission line model for hybridized polaritons. (a) An illustration of the dielectric/hBN/BLG/hBN/dielectric VdWH. (b) Transmission line schemes for the configuration in (a). (c) Dispersion relation of the hybridized polaritons in the configuration in the symmetric structure, (dashed purple line for even modes and dashed red line for odd modes) compared to TMM simulation (colormap). Even modes of HPhP for hBN with thickness of $d_b+d_t$ (black lines) and exciton energies of BLG (dashed white lines) are also plotted. (d) Dispersion relation of the hybridized polaritons in the configuration in the asymmetric structure, (dashed red line for the low momentum modes and dashed purple line for the high momentum modes) compared to TMM simulation (colormap). Modes of HPhP for hBN with thickness of $d_b+d_t$ (black lines), odd modes of HPhP for hBN with thickness of $2d_t$ (dashed orange lines), and exciton energies of BLG (dashed white lines) are also plotted.}
    \label{fig Hyb}
\end{figure}

To demonstrate the advantage of the TLM in the study of polaritons in VdWHs, we show the TLM analysis of a dielectric/hBN/BLG/hBN/dielectric VdWH, which supports both HPhPs and graphene excitons \cite{Eini2025Exciton--hyperbolic-phonon-polaritonGraphene} (Fig. \ref{fig Hyb} (a)), and where the physical meaning of the admittances plays an important role. The equivalent TLM of the VdWH is presented in Fig. \ref{fig Hyb} (b). The numerical simulation of the TMM  (colormap in Fig. \ref{fig Hyb} (c) and (d)) shows multiple modes, branches and anti-crossings. The derivation of an analytical expression of the dispersion relation and field distributions for such a complicated modal picture is very challenging via the TMM, or similar approaches, hindering the underlying physics of the system. Nevertheless, the TLM approach offers much simpler derivations that allow to understand the physical properties of the system. We start with the analysis of a symmetric structure with both hBN layers having a thickness of $d_t=d_b$. Due to the symmetry of the VdWH, it supports even and odd modes. Since the odd modes satisfy a nullification of the voltage at $z=0$, yielding no surface current on the bilayer graphene, the magnetic field is continuous and the odd modes are not affected by the presence of the BLG (dashed red line in Fig. \ref{fig Hyb} (c)). The even modes (dashed purple line) present an anti-crossing between the even modes of HPhPs (black line) and the exciton energies of the BLG (dashed white line), forming hybridized polaritons. The dispersion relation can be found using the TLM with the assumption that the conductivity of the BLG is much smaller than the characteristic admittance of the hBN, $|\sigma| \ll |Y_{hBN}|$ \cite{Eini2025Exciton--hyperbolic-phonon-polaritonGraphene} (Fig. \ref{fig Hyb} (c)). 

Next, we analyze an asymmetric structure with $d_t \ll d_b$, which supports two sets of modes with either low momentum and high momentum (Fig. (d)). The low momentum modes (dashed red line) present an anti-crossing between modes of HPhP (black line) and the exciton energies (dashed white line), since they satisfy $|\sigma| \ll |Y_{hBN}|$, and can be found using the TLM, similarly to the symmetric case. The high momentum modes, however, satisfy $|\sigma| \gg |Y_{hBN}|$, meaning that the graphene serves almost as a short circuit and therefore can be approximately described as a PEC, separating between the top and bottom hBN layers. Since $d_b$ is larger than all quantities with length dimensions, the bottom layer can be approximated as half infinite. We hence approximately receive odd modes of HPhPs with of thickness $2d_t$ (dashed orange line), where the more accurate dispersion relation can be derived using the TLM by keeping another order of $|\frac{Y_{hBN}}{\sigma}|$ \cite{Eini2025Exciton--hyperbolic-phonon-polaritonGraphene} (dashed purple line). Thus, the TLM approach enables to obtain analytical expressions of the hybridized polaritons in the VdWH, using the physical properties of the admittances in the system and the basic physical principle of symmetry.

\section{Field Distribution}
Next we show how the field distributions in the VdWHs can be easily obtained using the TLM method. The voltage in the $i$th layer is given by:
\begin{align}    \mathbb{E}_T(z)=\mathbb{E}^+_T(z_i)e^{i k_{i_z} (z-z_i)}[1+r(z_{i+1})e^{-i k_{i_z} (z-z_{i+1})}] ,
\end{align}
where $z_i$ and $z_{i+1}$ are arbitrary locations in the $i$th layer, could be chosen as the interfaces between the $i\mp1$th and $i$th layers for convenience, and $\mathbb{E}^+_T(z_i)$ is an amplitude defined as the forward propagating voltage at $z_i$. The reflection coefficients can be calculated using Eq. \ref{Eq ref}, where the characteristic and the $z$ dependent admittances are usually already calculated previously in order to find the total reflection coefficient or the dispersion relation of the polaritons in the VdWH. If there is an external impinging wave then it determines the amplitude in the first layer, while the amplitudes in the other layers can be found using the continuity of the voltage. If one is interested in finding the field distribution of the polaritons, then there is no impinging wave, and the field can be found up to an arbitrary constant. The current can be found directly from the voltage by multiplying it by $Y(z)$, or by multiplying each propagating wave in each layer separately by $\pm Y_i$ and the longitudinal field can be derived from the transverse field using Maxwell's equations.

\section*{Conclusions}
 In conclusion, we have shown that the interaction of VdWHs with light can be appropriately modeled using an equivalent TLM. We derived the TLM for bulk to monolayer 2D materials, composing VdWHs, obtaining reflection coefficients agreeing well with experimental results, and polariton dispersion relations with excellent agreement compared to numerical simulations. We used the TLM to gain physical insight on the optical and polaritonic response of VdWHs. This method can be used in future theoretical studies of interaction between light and VdWHs, enabling exploring and understanding physical phenomena of the interaction.

\section*{Methods}
The VdWHs investigated in this study were assembled via a dry transfer process, as detailed in Ref. \cite{Castellanos-Gomez2014DeterministicStamping}. Reflection measurements were performed by focusing a broadband light source onto the VdWHs through an objective lens. The reflected signal was collected by the same objective and then routed into a spectrometer for spectral analysis. The reflection signal $R$ was acquired from a region containing the complete hBN/TMD/hBN structure, whereas the reference signal $R_0$ was measured from an adjacent area in which the TMD layer was absent.

\section*{Acknowledgments}
I.E. acknowledges the Israeli Science Foundation personal grant number 865/24, the Ministry of Science and Technology grant number 0005757, and the support of the European Union (ERC, TOP-BLG, Project No. 101078192). Y. M. acknowledges the support of ISF Grant No. 1089/22. J.H.E. acknowledge support from the Office of Naval Research under award no. N00014-20-1-2474 for hBN crystal growth. S.T acknowledges primary support from DOE-SC0020653 (excitonic testing) and NSF CBET 2330110 (environmental stabilization), and support from Applied Materials Inc., and Lawrence Semiconductor Labs for material development / initial characterization.

\bibliography{TLM.bib}

\end{document}